\documentclass[10pt,letterpaper,twocolumn]{article} 

\usepackage{ol}
\usepackage{hyperref}
\usepackage{bm}
\usepackage{graphicx}
\usepackage{amsbsy}
\usepackage{amsmath}
\usepackage{amsfonts}

\newcommand{\bra}[1]{\langle{#1}|}
\newcommand{\ket}[1]{|{#1}\rangle}
\newcommand{\nn}{\nonumber \\}

\newcommand{\Tr}{{\rm Tr}}

\newcommand{\braket}[2]{\langle{#1}|{#2}\rangle}
\newcommand{\ketbra}[2]{|{#1}\rangle\langle{#2}|}

\newcommand{\s}{z}

\begin{document}\twocolumn[
\title{Interconvertibility of single-rail optical qubits}

\author{Dominic W. Berry} 
\affiliation{Department of Physics, The University of Queensland, Queensland
4072, Australia}
\author{Alexander I. Lvovsky and Barry C. Sanders}
\affiliation{Institute for Quantum Information Science, University of Calgary,
Alberta T2N 1N4, Canada}

\begin{abstract}
We show how to convert between partially coherent superpositions of a single
photon with the vacuum using linear optics and postselection based on homodyne
measurements. We introduce a generali\s ed quantum efficiency for such states
and show that any conversion that decreases this quantity is possible. We also
prove that our scheme is optimal by showing that no linear optical scheme with
generali\s ed conditional measurements, and with one single-rail qubit input
can improve the generali\s ed efficiency.
\end{abstract}

\ocis{270.5290, 270.5570}]

\date{\today}

\maketitle
The \emph{single-rail optical qubit} is a coherent superposition of the
single-photon and vacuum states of light: $\ket\psi=\alpha\ket 0+\beta\ket 1$.
Such qubits, along with their dual-rail siblings, are basic units of information
in quantum-optical information processing\cite{LundRalph,Ralph}.
Recently, several experiments implemented preparation of arbitrary single-rail
qubits from the single-photon state using linear optics and conditional
measurements. The quantum catalysis scheme\cite{catalysis} used an ancillary
coherent state input and conditional single-photon measurements. The quantum
scissors setup\cite{Pegg,Babichev} employed the Bennett quantum teleportation
protocol\cite{QTBennett} with a delocali\s ed single photon as an entangled
resource and a coherent state as an input. Most recently, a similar resource
was used for remote preparation\cite{RSP} of single-rail qubits via field
quadrature (homodyne) measurements by one of the entangled parties. 

\begin{figure}
\centerline{\includegraphics[width=2.9cm]{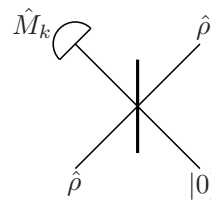}}
\caption{\label{fig:sing}The general scheme for converting one mixed state
of zero and one photon to another.}
\end{figure}

All these schemes can be reversed to generate a single-photon state, and, more
generally, another single-rail qubit from a single-rail qubit input. In this
paper, we concentrate on, and generali\s e, the setup of Babichev
{\it et al.}\cite{RSP}, which is shown in Fig.~\ref{fig:sing}. The initial
qubit, $\hat\rho$, is combined with vacuum at a beam splitter, generating a
two-mode state $\hat\rho_{\rm BS}$. A measurement described by a positive
operator-valued measure (POVM) is then performed on mode 1. Conditioned on
measurement result $k$, the output in mode 2 is
$\hat\rho'\propto{\rm Tr}_1(\hat M_k \hat\rho_{\rm BS})$.

For a pure input $\hat\rho=\ketbra{\psi}{\psi}$, and a projective measurement
in mode 1, the output $\hat\rho'$ is also a single-rail qubit. However, with
an imperfect input or generali\s ed measurement, the output state may not
be pure. In a realistic experiment, the single most significant
imperfection of the input state is the admixture of the vacuum\cite{RSP,Fock}:
\begin{equation}\label{rho}
\hat\rho = E\ket\psi\bra\psi +(1-E)\ket 0 \bra 0
\end{equation}
with $E\le1$ being the quantum efficiency (we take the convention that
$E=\beta=0$ for the vacuum state). The output state is then also of the
form (\ref{rho}), and we use primed symbols for the notation related to the
output state. In this Letter, we answer the following question: under which
circumstances can an imperfect single-rail qubit characteri\s ed by parameters
$(\alpha,\beta,E)$ be converted to another qubit with parameters
$(\alpha',\beta',E')$?

The initial state is transformed by the beam splitter of transmissivity $t^2$
and reflectivity $r^2$ to $\hat\rho_{\rm BS}=E\ket\phi\bra\phi+(1-E)\ket{00}$,
with $\ket\phi=\alpha\ket{00}+\beta r\ket{10}+\beta t\ket{01}$. We begin by
considering a projective measurement $\ketbra{Q}{Q}$. Projecting mode 1 onto a
state $\ket{Q}$ will yield the unnormali\s ed output state
\begin{align}
\hat\rho'&=E[(\alpha\theta_{0} +\beta r\theta_{1})\ket 0 +\beta t\theta_{0} \ket 1]
\otimes {\rm H.c.} \nn & \quad +(1-E) |\theta_{0}|^2 \ket 0\bra 0,
\end{align}
where $\theta_{j}=\braket{Q}{j}$, from which we find 
\begin{equation}\label{rrat}
\beta'(\alpha\theta_{0} +\beta r\theta_{1})=\alpha'\beta t\theta_{0}.
\end{equation}
The trace
\begin{equation}\label{Psucc}
\Tr[\hat\rho']=E(|\alpha\theta_{0} +\beta r\theta_{1}|^2+|\beta t\theta_{0}|^2)+(1-E) |\theta_{0}|^2
\end{equation}
is equal to (proportional to in the case of $Q$ being a continuous observable)
the probability for the desired measurement result to occur, and the efficiency
of the output qubit is 
\begin{equation}\label{Eprime}
E'=E(|\alpha\theta_{0} +\beta r\theta_{1}|^2+|\beta t\theta_{0}|^2)/\Tr[\hat\rho'].
\end{equation}
Using Eq.~(\ref{rrat}) as well as $\alpha^2+\beta^2=\alpha'^2+\beta'^2=1$, we
simplify Eqs.~(\ref{Psucc}) and (\ref{Eprime}) as follows:
\begin{align}
\label{tran}
t\beta \sqrt{E(1-E')} = \beta'\sqrt{E'(1-E)}, \\
p_{\rm success}\equiv\Tr[\hat\rho']=|\theta_0|^2(1-E)/(1-E').
\end{align}

Eq.~(\ref{rrat}) determines the efficiency-independent parameters of the output
qubit that can be controlled by choosing the beam splitter and the measurement.
An example is a field quadrature measurement using a homodyne detector\cite{RSP}.
With local oscillator phase $\phi$, result $Q$ gives the projection
$\bra{Q_\phi}$ satisfying (scaling convention is $[\hat Q,\hat P]=i/2$)
\begin{align}
\theta_0=\braket {Q_\phi}0 &= \left( 2/\pi\right)^{1/4}e^{-|Q|^2}, \nn
\theta_1=\braket {Q_\phi}1 &= 2Q\left( 2/\pi\right)^{1/4}e^{-|Q|^2}e^{i\phi}, 
\end{align}
so $\theta_1/\theta_0=2Qe^{i\phi}$.
With any beam splitter, by choosing
appropriate $Q$ and $\phi$, one can obtain any desired qubit transformation
except the transformation to vacuum. (The trivial transformation to vacuum may
be obtained by taking $t=0$.)

We now turn to restrictions on the efficiency of the output qubit. 
We define the {\it generali\s ed} efficiency by
\begin{equation}
\mathcal{E}(\hat\rho) \equiv \frac{\rho_{11}}{1-|\rho_{01}|^2/\rho_{11}}
=\frac{|\beta|^2E}{1-|\alpha|^2E}, 
\end{equation}
where $\rho_{ij}=\bra i\hat\rho\ket j$, and the $\rho_{ij}$ are given by
\begin{equation}\label{rhomatrix}
(\rho_{ij})=\left[\begin{array}{*{10}c}
1-E|\beta|^2 & E\alpha^*\beta \\
E\alpha\beta^* & E|\beta|^2\end{array}\right].
\end{equation}
This efficiency has a number of useful properties: it is convex (see Appendix
A); it reduces to $E$ for imperfect single photon sources ($\alpha=0$); $\mathcal{E}(\hat\rho)=1$ 
corresponds to pure states, and
$\mathcal{E}(\hat\rho)=0$ corresponds to the vacuum state.
Most importantly, since the beam splitter transmission cannot exceed one, Eq.~(\ref{tran}) entails 
$\mathcal{E}(\hat\rho)\ge\mathcal{E}(\hat\rho')$.
Hence \emph{only those transformations that do not increase the generali\s ed efficiency are possible}. 

Furthermore (see Appendix B), \emph{our scheme with homodyne measurements
permits any transformation that decreases the generali\s ed efficiency}.
There is a complication for the case $\mathcal{E}(\hat\rho)=1$ and
$\mathcal{E}(\hat\rho)>\mathcal{E}(\hat\rho')$: a mixed state cannot be
generated from a pure one using a projective measurement. In this case
we can slightly attenuate the input state so its generali\s ed efficiency
reduces to a value between 1 and $\mathcal{E}(\hat\rho')$. We can
then transform the state to $\hat\rho'$ using a second beam splitter and
projective measurement. Mathematically, this procedure is equivalent to that of
Fig.\ \ref{fig:sing} with a POVM.

There are also two cases where it is possible to transform between
states of equal generali\s ed efficiency. These cases are: \\
Case 1: $\mathcal{E}(\hat\rho)=\mathcal{E}(\hat\rho')=1$ \\
Case 2: $E=E'$ and $|\beta|=|\beta'|$\\
Case 1 corresponds to transforming between different pure states. Case 2
corresponds to a trivial phase shift, and also includes the trivial vacuum to
vacuum transform. One can show (Appendix B) that these cases are the only ones
where a transform between states of equal efficiency is possible.

\begin{figure}
\centerline{\includegraphics[width=8.64cm]{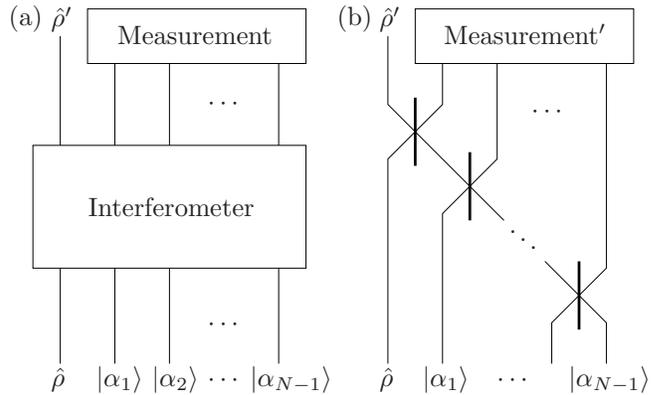}}
\caption{\label{fig:abs} (a) A general setup for transforming the state
$\hat\rho$ to $\hat\rho'$ involves an interferometer and a conditional measurement.
(b) The same interferometer after absorbing all but a line of beam splitters into the
measurement.}
\end{figure}

We now generali\s e the obtained efficiency enhancement restriction to any
scheme involving linear optics and conditional measurements
(Fig.~\ref{fig:abs}(a)). We assume that, first, there is only one nonclassical
input: the single-rail qubit $\hat\rho$ and, second, the output state
$\hat\rho'$ has only zero- and one-photon terms in its Fock decomposition.
Consider a processing scheme where the state is combined with any
number of vacuum and coherent states at the input, then a POVM measurement
is performed on all the modes except 1. A $U(N)$ interferometer can
be decomposed into a line of beam splitters followed by a $U(N-1)$
interferometer\cite{berry2}. The $U(N-1)$ interferometer can then be absorbed into
the measurement, so the resulting interferometer is as in Fig.\ \ref{fig:abs}(b).

The beam splitters leave the coherent modes unentangled. One of these is an
input to the last beam splitter, whereas a measurement is performed
on the others. As these modes are not entangled, they may be omitted entirely,
so there is only one coherent state input. In order to eliminate the multiphoton
components in $\hat\rho'$, the amplitude of this coherent state must be zero.
The simplified configuration is then just as in Fig.\ \ref{fig:sing}. 

We now recall that the measurement on mode 2 is generally not a projective
measurement, but a POVM measurement. However, by a singular value decomposition,
$\hat M_k = \sum p_i\ket{\sigma_i}\bra{Q_i}$, where the $\ket{Q_i}$'s are
orthogonal. The state $\hat\rho'$ is thus a statistical mixture of outputs
$\hat\rho'_i$ associated with projections onto $\ket{Q_i}$. Because the
generali\s ed efficiency is convex (see Appendix A), we find $\forall i$
$\mathcal{E}(\hat\rho')\le\mathcal{E}(\hat\rho'_i)\le\mathcal{E}(\hat\rho)$,
which completes the proof.

One application of our general scheme is to transform an
incoherent mixture of the single-photon and vacuum states to a partially
coherent state; this corresponds to the state preparation achieved by
Refs.~\citenum{catalysis,Babichev,RSP}. A reverse transformation is also
possible. In other words, the state $\hat\rho$ is equivalent to a photon source
with efficiency $\mathcal{E}(\hat\rho)$, in the sense that we may interconvert
between these two states arbitrarily accurately. Therefore, it is reasonable to
state that the generali\s ed efficiency we have defined for mixed states with
partial coherence is equivalent to the usual efficiency for single photon
sources\cite{Fock}. 

Whereas the generali\s ed efficiency cannot be increased, it is possible to
enhance the specific efficiency $E$ of a state, as demonstrated
experimentally\cite{RSP}. However, this will occur at the expense of the
single-photon fraction $\beta$ in the qubit part of the state. It is not
possible, for example, to enhance the efficiency of a single-photon source by
first converting it into a single-rail qubit and then back into a single-photon
state: while the efficiency may increase in the first step, it will reduce in
the second step to no more than its original value.

Recently, Berry {\it et al.} proved for some special cases that it is impossible
to enhance the efficiency of a single-photon source with any interferometric
scheme such as in Fig.~\ref{fig:abs}(a) but with an arbitrary number of inefficient
single-photon inputs\cite{efficiencyimprovement}. If this limitation is correct
in general, the impossibility proof made in this paper would also be valid to
the same degree of generality. Indeed, if there existed a scheme allowing one to
conditionally convert an inefficient single-rail qubit $\hat\rho$ to another
$\hat\rho'$ so that $\mathcal{E}(\hat\rho')>\mathcal{E}(\hat\rho)$, one could
use it to enhance the efficiency of a single-photon source. One would first set
up a circuit as in Fig.\ \ref{fig:sing} that converted the single photon to
$\hat\rho$ with small loss of generali\s ed efficiency, then transform
$\hat\rho$ to $\hat\rho'$ enhancing the generali\s ed efficiency, then convert
$\hat\rho'$ back to a single-photon state.

In summary, we have presented a general state transformation scheme for
single-rail qubits. This scheme uses only a beam splitter and conditional
measurement, and includes existing experimental arrangements as special cases.
We have also introduced a generali\s ed measure of the
efficiency of partially coherent mixed states of zero and one photon. This
efficiency corresponds to the efficiency of imperfect single photon sources,
because it is possible to reversibly convert between this state and a single
photon source with arbitrarily low efficiency loss.
Our scheme allows any transformation that decreases this generali\s ed efficiency.
Transformations that increase the efficiency are not possible, and
transformations that keep the generali\s ed efficiency constant are not possible,
except for some trivial cases.

\textit{Appendix A: Convexity Proof. -- }
Here we show that
\begin{equation}
\label{eq:con}
\mathcal{E}(p\hat\rho_1+(1-p)\hat\rho_2)\le p \mathcal{E}(\hat\rho_1)+
(1-p) \mathcal{E}(\hat\rho_2) ,
\end{equation}
for $p\in[0,1]$. If both $\hat\rho_1$ and $\hat\rho_2$ are the vacuum state, then
clearly equality is achieved. If one of $\hat\rho_1$ and $\hat\rho_2$ is the vacuum
state, then we can take that state to be $\hat\rho_2$ without loss of generality.
Then, according to Eq.~(\ref{rhomatrix}),
\begin{align}
\mathcal{E}(p\hat\rho_1+(1-p)\hat\rho_2) &= \frac{p|\beta|^2E}{1-p|\alpha|^2E}
\le \frac{p|\beta_1|^2E_1}{1-|\alpha_1|^2E_1} \nn
&=p \mathcal{E}(\hat\rho_1)+(1-p) \mathcal{E}(\hat\rho_2).
\end{align}

For the case where neither state is the vacuum we define the function
$f(p)=\mathcal{E}[p\hat\rho_1+(1-p)\hat\rho_2]$. Taking the second derivative of this
function, we obtain
\begin{align}
f''(p)&=\frac{2\rho_{11}^2|\Delta\rho_{10}|^2}
{(\rho_{11}-|\rho_{10}|^2)^2} \nn
&\quad +\frac{2[\Delta\rho_{11}|\rho_{10}|^2
-\rho_{11}2{\rm Re}(\rho_{01}\Delta\rho_{10})]^2}
{(\rho_{11}-|\rho_{10}|^2)^3},
\end{align}
where $\rho_{ij} = \bra i [p\hat\rho_1+(1-p)\hat\rho_2] \ket j$ and
$\Delta\rho_{ij}=\bra i (\hat\rho_1-\hat\rho_2) \ket j$.
As neither state is the vacuum, this second derivative exists and is
non-negative for $p\in[0,1]$. Hence $f(p)$ is convex, which implies
Eq.\ \eqref{eq:con}. Thus, in each case we find that Eq.\ \eqref{eq:con} holds,
so the generali\s ed efficiency is convex.

\textit{Appendix B: Optimality Proof. --}
We first show that transforms that decrease the generali\s ed entropy are
possible. If $\mathcal{E}(\hat\rho)>\mathcal{E}(\hat\rho')$, then
$|\beta|^2E(1-E')>|\beta'|^2E'(1-E)$, so there is a solution of
Eq.\ \eqref{tran} with $|t|<1$ and $r\ne 0$. If $\beta'=0$, the transform can
be obtained with $t=0$. If $\beta'\ne 0$, then from Eq.\ \eqref{rrat} $\theta_{0}$ is also
nonzero. The probability for success is then nonzero for $E<1$.

Now we show that the transformation that preserves the generali\s ed
efficiency is possible for Case 1 (Case 2 is trivial).
For Case 1, $E=E'=1$, so Eq.\ \eqref{tran} is satisfied regardless of the value
of $t$. Taking $t=r=1/\sqrt 2$, $r\beta\beta'\ne 0$, so $\theta_{0}\ne 0$. The
probability of success is then $|\theta_{0}|^2(|t\alpha'\beta/\beta'|^2+
|\beta t|^2)$, which is nonzero, so the transformation is possible.

If $\mathcal{E}(\hat\rho)=\mathcal{E}(\hat\rho')$ but Cases 1 and 2 do not hold, then
$|\beta|^2E(1-E')=|\beta'|^2E'(1-E)$. For a projective measurement we obtain
$|t|=1$, so $r=0$ and $t\alpha'\beta\ne\alpha\beta'$. Then Eq.\ \eqref{rrat}
gives $\theta_{0}=0$, so the probability for success is zero.


\begin{thebibliography}{99}
\bibitem{LundRalph} A. P. Lund and T. C. Ralph, Phys. Rev. A {\bf 66}, 032307
(2002).
\bibitem{Ralph} T. C. Ralph, A. P. Lund, and H. M. Wiseman, quant-ph/0507192
(2005).
\bibitem{catalysis}  A. I. Lvovsky and J. Mlynek, Phys. Rev. Lett. {\bf 88},
250401 (2002).
\bibitem{Pegg} D. T. Pegg, L. S. Phillips, and S. M. Barnett,
\prl {\bf 81}, 1604 (1998).
\bibitem{Babichev} S. A. Babichev, J. Ries, and A. I. Lvovsky, Europhys. Lett.
{\bf 64}, 1 (2003).
\bibitem{QTBennett} C. H. Bennett, G. Brassard, C. Cr\'epeau, R. Jozsa,
A. Peres, and W. K. Wootters, Phys. Rev. Lett. {\bf 70}, 1895 (1993).
\bibitem{RSP} S. A. Babichev, B. Brezger, and A. I. Lvovsky, Phys. Rev. Lett.
{\bf 92}, 047903 (2004).
\bibitem{Fock}  A. I. Lvovsky, H. Hansen, T. Aichele, O. Benson, J. Mlynek, and
S. Schiller, Phys. Rev. Lett. {\bf 87}, 050402 (2001).
\bibitem{berry2} D. W. Berry, S. Scheel, C. R. Myers, B. C. Sanders,
P. L. Knight, and R. Laflamme, New J. Phys. {\bf 6}, 93 (2004).
\bibitem{efficiencyimprovement} D. W. Berry, S. Scheel, B. C. Sanders, and
P. L. Knight, Phys. Rev. A {\bf 69}, 031806 (2004).
\end{thebibliography}
\end{document}